\documentclass[%
 reprint,
superscriptaddress,
 amsmath,amssymb,
 aps,
pra,
floatfix,
]{revtex4-2}

\usepackage{xcolor}
\usepackage{svg}
\usepackage{multirow}
\usepackage{braket}
\usepackage{adjustbox}
\usepackage{makecell}
\usepackage{lipsum}
\usepackage{tabularray}
\usepackage{chngcntr}
\usepackage{longtable}
\usepackage{placeins}

\usepackage{graphicx}
\usepackage{dcolumn}
\usepackage{bm}
\usepackage[colorlinks = true,
            linkcolor = blue,
            urlcolor  = blue,
            citecolor = blue,
            anchorcolor = blue]{hyperref}

\usepackage{txfonts}

\usepackage{soul}
\usepackage{bbold}

\newcommand{\comment}[1]{}
\newcommand{\eref}[1]{Eq.~\eqref{#1}}
\newcommand{\sref}[1]{sec.~\ref{#1}}
\newcommand{\fref}[1]{Fig.~\ref{#1}}

\def \SigmaCon {\Sigma^c}
\def \SigmaUn {\Sigma^u}
\def \XiEx {\Xi^{\rm ex}}

\def \PEPR {\mathbf{P}_{\rm EPR}}

\graphicspath{ {Figures/} }

\begin{document}


\title{Steady-state entanglement of interacting masses in free space through optimal feedback control}

\author{Klemens Winkler}
    \email{klemens.winkler@univie.ac.at}
    \affiliation{University of Vienna, Faculty of Physics, Vienna Center for Quantum Science and Technology (VCQ), Boltzmanngasse 5, A-1090 Vienna, Austria}

\author{Anton V. Zasedatelev}
\email{anton.zasedatelev@aalto.fi}
\affiliation{University of Vienna, Faculty of Physics, Vienna Center for Quantum Science and Technology (VCQ), Boltzmanngasse 5, A-1090 Vienna, Austria}
\affiliation{Department of Applied Physics, Aalto University School of Science, P.O. Box 15100, Aalto FI-00076, Finland}

\author{Benjamin A. Stickler}

    \affiliation{Institute for Complex Quantum Systems, Ulm University, Albert-Einstein-Allee 11, D-89069 Ulm, Germany}

\author{Uroš Delić}

       \affiliation{University of Vienna, Faculty of Physics, Vienna Center for Quantum Science and Technology (VCQ), Boltzmanngasse 5, A-1090 Vienna, Austria}
       
\author{Andreas Deutschmann-Olek}

    \affiliation{Automation and Control Institute, TU Wien, A-1040, Vienna, Austria}

\author{Markus Aspelmeyer}
\email{markus.aspelmeyer@univie.ac.at}
    \affiliation{University of Vienna, Faculty of Physics, Vienna Center for Quantum Science and Technology (VCQ), Boltzmanngasse 5, A-1090 Vienna, Austria}
    \affiliation{Institute for Quantum Optics and Quantum Information (IQOQI) Vienna, Austrian Academy of Sciences, Boltzmanngasse 3, A-1090 Vienna, Austria}

\date{\today}

\begin{abstract}
We develop a feedback strategy based on optimal quantum feedback control for Gaussian systems to maximise the likelihood of steady-state entanglement detection between two directly interacting masses. We employ linear quadratic Gaussian (LQG) control to engineer the phase space dynamics of the two masses and propose Einstein-Podolsky-Rosen (EPR)-type variance minimisation constraints for the feedback to facilitate unconditional entanglement generation. This scheme allows for stationary entanglement in parameter regimes where strategies based on total energy minimisation (\textit{cooling}) would fail. 

\end{abstract}

\maketitle

\section{Introduction}

The generation of entanglement in the motion of two large masses has recently been achieved through cavity-mediated optomechanical interactions~\cite{Riedinger_2018, Ockeloen_2018,Wollack_2022}. In contrast, entanglement through direct electromagnetic (Coulomb) or gravitational coupling remains elusive, and yet it holds conceptual importance in exploring the quantum nature of the fields underlying the entanglement generation~\cite{dewitt2011role, Belenchia2018QuantumSuperposition, Bose_2017, Marletto_2017, Christodoulou_2023, Martinez_2023}. 
The challenge arises from the large decoherence rates of quantum states in macroscopic systems~\cite{RevModPhys.90.025004}, which must not exceed the coupling rates for successful entanglement generation, whether these interactions are induced via weak gravitational, Casimir, or stronger Coulomb forces. The latter, electrostatic case is of particular interest, as it offers realistically large coupling rates and is similar in nature to the gravitational interaction with its $1/r$ potential~\cite{Belenchia2018QuantumSuperposition}.
Levitated solids in ultra-high vacuum provide an excellent platform to achieve entanglement between large-mass systems through direct physical interactions~\cite{Ballestero_2021}. Recent theoretical efforts have explored this problem in both steady-state~\cite{Rudolph_2022} and dynamical realms~\cite{PhysRevLett.127.023601, Cosco_2021}. Even under favourable conditions, such as strong Coulomb interactions between optically trapped particles \cite{Rieser_2022, Deplano_2024} and state-of-the-art experimental parameters for motional ground-state cooling~\cite{Magrini_2021, Tebbenjohanns_2021}, the interaction strength $g$ must be extremely high, \( |g|/\Omega_0 \sim 2 \), to achieve unconditional entanglement~\cite{Rudolph_2022}.

To facilitate entanglement generation, methods of optimal feedback control theory have recently been adopted for optomechanical systems~\cite{Hofer_2015, Miki_2023}. In particular, the linear quadratic Gaussian (LQG) control strategy, which combines optimal state estimation and feedback control, has been extensively studied both theoretically~\cite{Wiseman_2009, Hofer_2015} and experimentally~\cite{Wieczorek_2015, Setter_2018, Magrini_2021}. This strategy has emerged as a promising approach for achieving quantum control in various optomechanical systems.
In this work, we study different feedback configurations along with optimization objectives of LQG control to maximize the likelihood of entanglement generation. As a result, using an optimized Einstein-Podolsky-Rosen (EPR)-type constraint, we achieve unconditional entanglement generation under repulsive electrostatic interactions with an order of magnitude lower interaction strength (\(|g|/\Omega_0 \sim 0.2\)) compared to the regime of attractive interactions. Furthermore, as is shown in \cite{poddubny2024nonequilibrium}, when employed in systems with time-dependent coupling, our strategy can push unconditional entanglement to its fundamental quantum limit.

This article is structured as follows: In \sref{sec:Sys_dynm} we introduce the  interaction and dynamics of the bare and normal mechanical modes. We present the measurement and feedback mechanisms, as well as entanglement criteria used throughout this work. In \sref{sec:QuantumControl} we provide the formalism for  quantum optimal control  based on Kalman filter and linear quadratic regulator, together forming a LQG control problem. In \sref{sec:Results}, we present the results for entanglement generation in conditional and unconditional states using different  feedback configurations and cost-functions.

\begin{figure}
    \centering
    \includegraphics[width=\columnwidth]{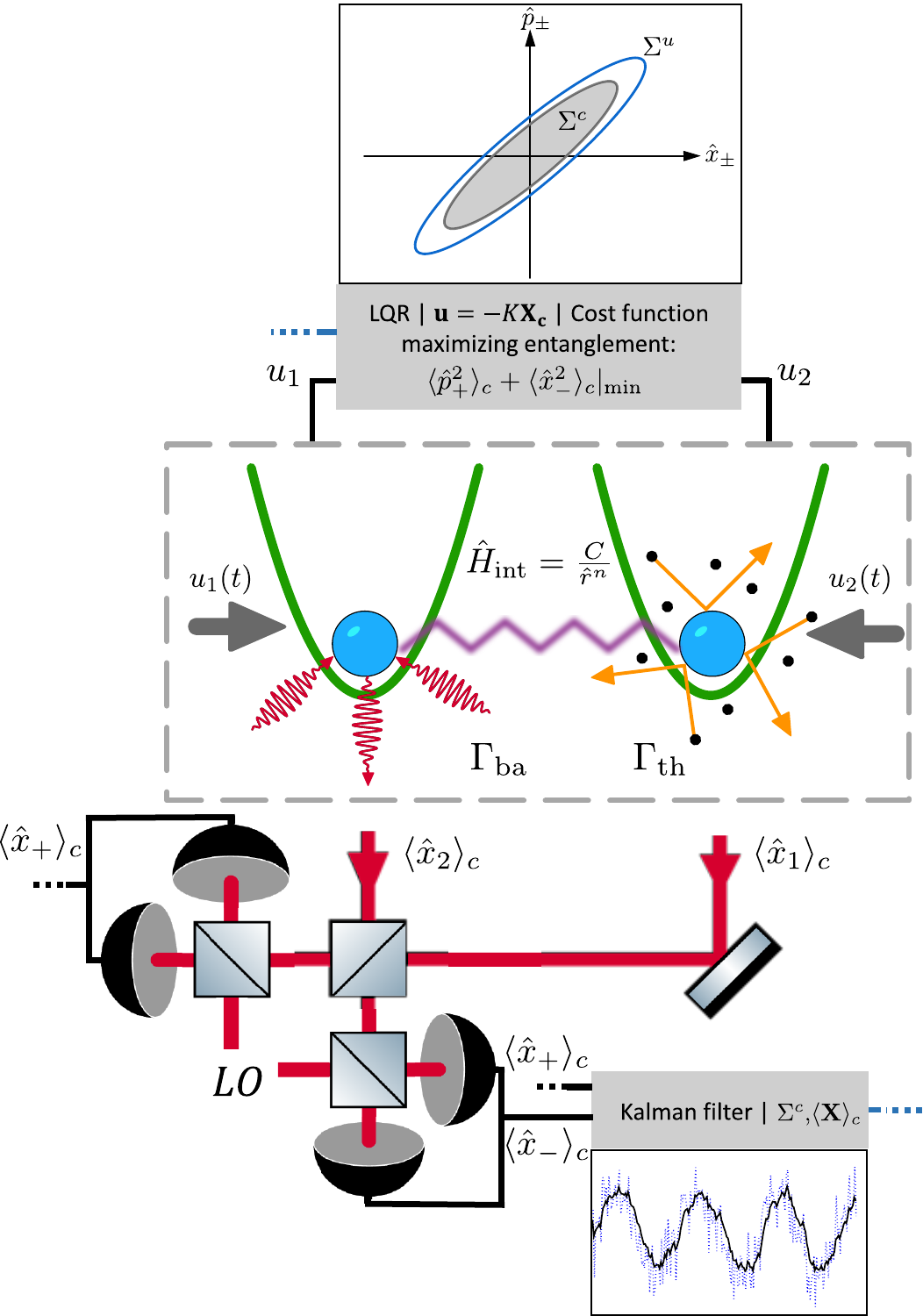}
   \caption{\label{fig:Sketch} Two particles trapped in different harmonic traps interacting via a $1/r^{n}$-potential. Both particles are subject to thermal decoherence due to random kicks from residual gas with decoherence rate $\Gamma_{\rm th}$ (right). Optical photons are scattered from the particles (left), introducing back-action noise with decoherence rate $\Gamma_{\rm ba}$ while carrying information on the position of each particle. The back-scattered photons are independently collected, mixed on a balanced beamsplitter and measured via homodyne detection with efficiency $\eta$, accounting for any losses of photons. From the measurement records, the Kalman filter estimates the conditional means $\mathbf{X}_c$ and covariances $\SigmaCon$ which are used by the linear quadratic regulator (LQR) for the optimal feedback input $\mathbf{u}=(u_1, u_2)^T$. Both feedback inputs (grey arrows) are applied to each particle, closing the feedback-loop.} 
\end{figure}
\section{Systems Dynamics}\label{sec:Sys_dynm}

\subsection{Bare modes dynamics}\label{sec:bare_modes}

In the following, we assume a  setup of two identical particles with mass $m$ trapped in a one-dimensional harmonic potential with a mechanical frequency $\Omega_0$ and motion $\hat{X}_{1,2}$, being separated by a mean distance $d$ in an orthogonal direction. 
We assume the particles to be coupled via a central interaction with Hamiltonian in the form: $\hat{H}_{\rm int}=C/\hat{r}^n$, where $\hat{r}= \sqrt{(\hat{X}_1-\hat{X}_2)^2+d^2}$ and $C$ is the interaction constant. 
For small particle displacements compared to $d$ we can Taylor-expand the interaction Hamiltonian to a second order in $(\hat{X}_1-\hat{X}_2)$, which results in the quadratic interaction term $\hat{H}_{\rm int}\approx -C(\hat{X}_1-\hat{X}_2)^2/\left(2 d^{2+n} n \right)$ \footnote{Note that we dropped  constant terms since they do not contribute to the particles dynamics}. Introducing the dimensionless position quadrature $\hat{x}_k = \hat{X}_k/x_{\rm zpf}$ with the zero-point fluctuation $x_{\rm zpf}= \sqrt{\hbar /m \Omega_0}$, we define the coupling rate
\begin{equation}\label{eq:coupling_rate}
    g = -\frac{C x_{\rm zpf}^2}{2 \hbar d^{2+n} n }. 
\end{equation}
For an attractive (repulsive) interaction, where $C<0$ $(C>0)$, we have $g>0$ $(g<0)$. 
To control the motion of both particles, we assume them to carry excess charges $Q_{1,2}$. This allows the mechanical system  to be steered by a time-dependent electric field, given the field strength can be controlled at the position of each particles, that gives rise to  feedback terms $\hat{H}^{k}_{\rm fb} = -Q_k E^{k}_{\rm fb}(t) x_{\rm zpf} \hat{x}_k= \hbar u_k(t) \hat{x}_k$ in the total Hamiltonian of the system: 

\begin{equation}\label{eq:Hamiltonian}
    \hat{H} = \sum_{k=1,2}\frac{\hbar \Omega_0}{2}\left(\hat{x}_k^2+ \hat{p}_k^2\right) -\sum_{k=1,2}\hbar u_k(t)\hat{x}_k+\hbar g \left(\hat{x}_1- \hat{x}_2\right)^2 
\end{equation} 
formulated for the dimensionless mechanical position and momentum quadratures $\hat{x}_k$ and $\hat{p}_k = \hat{P}_k/p_{\rm zpf}$ with $p_{\rm zpf}=\sqrt{\hbar m\Omega_0}$, where $k \in \{1,2\}$.

Both particles experience decoherence due to gas molecule scattering, and back-action from continuous position measurement by photon scattering off the particles into free space \cite{Tebbenjohanns_2019}.
We assume both processes to be Markovian and independent with respect to each other and quantify their respective strength by the thermal and back-scattering decoherence rates $\Gamma_{\rm th}$ and $\Gamma_{\rm ba}$. 

The light back-scattered from the particles is measured by two independent homodyne detectors. The information obtained by this continuous measurement is then used to construct the conditional state of the joint system represented by the density matrix $\hat{\rho}_c(t)$. The corresponding master equation is given by \cite{Gardiner_2010, Jacobs_2014}

\begin{equation}\label{eq:master_cond}
    \begin{gathered}
        d\hat{\rho}_c= -\frac{i}{\hbar} \left[\hat{H}, \hat{\rho}_c(t)\right]dt -\sum_{k=1}^2 \mathcal{D}_{\rm cl}^k[\hat{\rho}_c]dt \\
        -\sum_{k=1}^2\mathcal{D}_{\rm pr}^k[\hat{\rho}_c]dt+\sum_{k=1}^2 \mathcal{H}^k[\hat{\rho}_c] dW_k
    \end{gathered}
\end{equation}
where we introduced the Caldeira-Leggett type dissipator $\mathcal{D}_{\rm cl}^k$ \cite{Caldeira_1983}, as well as dissipation due to photon recoil $\mathcal{D}_{\rm pr}^k$ \cite{Henning_2021_b}. The last term of \eref{eq:master_cond}  given by the superoperator $\mathcal{H}^k$ accounts for the conditioning on the homodyne measurement with the classical Wiener increment $\mathbb{E}[d W_k]=0$, $\mathbb{E}[d W_k d W_l]=\delta_{kl}dt/2$. 
Introducing the mechanical damping rate $\gamma$, connected to the thermal decoherence rate via the mean thermal occupation number $\Gamma_{\rm th}=\gamma \bar{n}$, and the measurement rate $\Gamma_m$ representing an efficiency of motion readout by the homodyne detection ($\eta = \Gamma_{\rm m}/\Gamma_{\rm ba} $)~\cite{Magrini_2021}, we define superoperators as follows

\begin{subequations}
    \begin{gather}
        \mathcal{D}_{\rm cl}^k[\hat{\rho}_c]:= \frac{i\gamma}{2}\left[\hat{x}_k, \{\hat{p}_k, \hat{\rho}_c\}\right]+\Gamma_{\rm th}\left[\hat{x}_k, \left[\hat{x}_k, \hat{\rho}_c\right]\right]\label{eq:Caldeira}\\
        \mathcal{D}_{\rm pr}^k[\hat{\rho}_c]:= \Gamma_{\rm ba}\left[\hat{x}_k, \left[\hat{x}_k, \hat{\rho}_c\right]\right]\\
        \mathcal{H}^k[\hat{\rho}_c]:= \sqrt{4\Gamma_{\rm m}}\{\hat{x}_k-\langle\hat{x}_k\rangle_c , \hat{\rho}_c\}.
    \end{gather}
\end{subequations}  
where we use the following notation $\langle \hat{O}(t)\rangle_c=Tr[\hat{O}\hat{\rho}_c(t)]$. We note that the dissipator in \eref{eq:Caldeira} in general does not preserve  positivity of the density matrix $\hat{\rho}_c$. However, for  damping rates $\gamma\ll g, \Omega_0, \Gamma_{\rm th},\Gamma_{\rm ba}$, as considered later in this work, the states remain physical. 


By measuring the position quadrature via two independent homodyne detectors we effectively condition the motional states of both particles described by the master equation above \eref{eq:master_cond} to the measurement outcome in the form of a photocurrent at the detectors $I_k(t)$~\cite{Hofer_2015}

\begin{equation}\label{eq:meas_current_bare}
I_k(t) dt= \sqrt{4\Gamma_m}\langle \hat{x}_k\rangle_c dt + dW_k(t).
\end{equation}

\subsection{Normal mode transformation}\label{sec:normal_modes}

As evident from the Hamiltonian in \eref{eq:Hamiltonian} the interaction $\hbar g\left(\hat{x}_1-\hat{x}_2\right)^2$ couples the motion of both particles.  
Nevertheless, introducing normal modes through a symplectic transformation can effectively decouple the time evolution of the particles
\begin{equation}\label{eq:transform}
    \hat{x}_{\pm} = \alpha_{\pm}\frac{\hat{x}_1\pm\hat{x}_2}{\sqrt{2}} \quad \hat{p}_{\pm} = \frac{1}{\alpha_\pm}\frac{\hat{p}_1\pm\hat{p}_2}{\sqrt{2}}
\end{equation}
where $\alpha_\pm=\sqrt{\Omega_\pm/\Omega_0}$. This transformation preserves the commutation relations of the normal modes, i.e. $[\hat{x}_\pm, \hat{p}_\pm]=[\hat{x}_{1,2}, \hat{p}_{1,2}]=i$. Here, we introduced the eigenfrequenies of the common and differential modes as $\Omega_+ = \Omega_0$ and $\Omega_- =  \sqrt{\Omega_0^2+4g\Omega_0}$. Depending on the sign of the coupling rate $g$, the differential eigenfrequency will be greater or lower than $\Omega_0$ for attractive ($g>0$) and repulsive ($g<0$) interaction, respectively.
To maintain the stability of the mechanical system, it is necessary to keep the repulsive interaction to coupling rates within $0>g/\Omega_0 >-0.25$. Increasing the repulsive interaction toward $g = -0.25$ effectively reduces the mechanical stiffness for the differential mode, such that its frequency goes down to zero. Practically, the particles escape the trapping potential when the differential mode becomes unstable. 

Although the symplectic transformation modifies the thermal and back-action noise for the normal modes, it still preserves their uncorrelated nature, as shown in \eref{eq:Vmat}. For the following discussion we define the feedback forces for both normal modes 
\begin{equation}
     u_\pm(t) = \frac{1}{\alpha_{\pm}}\frac{u_1(t) \pm u_2(t)}{\sqrt{2}}.
\end{equation}

It is convenient to directly readout the motion of normal modes in the homodyne detection, for example, by implementing the transformation through mixing two independent optical detection channels, as shown schematically in \fref{fig:Sketch}. Hereinafter, we assume that the back-scattered light from each particle is mixed on a $50/50$ beam-splitter where each of the outputs is then measured independently. The resulting photocurrent $I_\pm(t)$  therefore exhibits linear dependence on the mean value of position quadratures of the normal modes, similar to the photocurrent in the separate detection of particles \eref{eq:meas_current_bare}: 

\begin{equation}\label{eq:meas_current_normal}
I_\pm(t) dt= \frac{\sqrt{4\Gamma_m}}{\alpha_\pm}\langle \hat{x}_\pm\rangle_c dt + dW_\pm(t)
\end{equation}
where $dW_\pm$ represents the classical zero-mean Wiener increment $\mathbb{E}[dW_i dW_j]= \delta_{ij}dt/2$ with $i,j \in \{+,-\}$, which are identical to the increments introduced in \eref{eq:meas_current_bare} for the separate detection. 

Assuming a Gaussian initial state for the system, the Hamiltonian in \eref{eq:Hamiltonian}, being at most quadratic in the mechanical quadratures, together with the master equations from \eref{eq:master_cond} and the homodyne photocurrent in \eref{eq:meas_current_normal}, ensure that the conditional state remains Gaussian throughout its time evolution \cite{Wiseman_2009, Rudolph_2022}. Thus the conditional state of the system, including its entanglement properties, are fully determined by the corresponding first and second moments.  

We collect all normal mode quadratures in a single vector $\hat{\mathbf{X}}=\left(\hat{x}_+,\hat{p}_+,\hat{x}_-,\hat{p}_-\right)^T$ and introduce the conditional mean vector $\mathbf{X}_c = \langle \mathbf{X}\rangle_c$
together with the respective real $4\times 4$ conditional covariance matrix $\Sigma_{kl}^c = \mathrm{Re}\left(\langle X_k X_l\rangle_c\right)-\langle X_k \rangle_c\langle X_l \rangle_c$ with $k, l \in \{+,-\}$. 
With  the real $4\times4$ drift matrix $\mathbf{A}$, real $4\times2$ control matrix $\mathbf{B}$, input force vector $\mathbf{u}$ of dimension $2$, the real $2\times4$ measurement matrix $\mathbf{C}$, the vector $d\mathbf{w}(t) = \left( dW_+(t),dW_-(t) \right)^T$, the  noise correlations matrices of  system noise $\mathbf{V}$,  measurement noise $\mathbf{W}$ and  cross-correlation between system and measurement noise $\mathbf{M}$ as
\begin{subequations}\label{eq:normal_matform}
    \begin{gather}
        \mathbf{A} = \begin{pmatrix}
            \mathbf{A}_+ & 0\\
            0 & \mathbf{A}_-
        \end{pmatrix} \quad \mathbf{A}_\pm = 
        \begin{pmatrix}
            0 & \Omega_\pm \\
            -\Omega_\pm & -\gamma
        \end{pmatrix}\\
        \mathbf{B} = \begin{pmatrix}
            0 & 0 & 0 & 1\\
            0 & 1 & 0 & 0
        \end{pmatrix}^T \label{eq:control_matrix}\\
           \mathbf{u}(t) = \left(u_+(t), u_-(t)\right)^T\\
           \mathbf{C} =
          \sqrt{4\Gamma_{\rm m}} \begin{pmatrix}
               \frac{1}{\alpha_+} & 0 & 0 & 0\\
               0 & 0 & \frac{1}{\alpha_-} & 0 
           \end{pmatrix}\label{eq:measurement-mat}\\
           \mathbf{V} = \text{\rm diag}\left(0, \frac{2(\Gamma_{\rm ba}+\Gamma_{\rm th})}{\alpha_+^2}, 0,\frac{2(\Gamma_{\rm ba}+\Gamma_{\rm th})}{\alpha_-^2}\right)\label{eq:Vmat}\\
       \mathbf{W} = \frac{1}{2}\mathbb{1}_{2\times 2}\\
       \mathbf{M}= 0
    \end{gather}
\end{subequations}
we find for the time evolution of the conditional means and covariances \cite{Edwards_2005, Hofer_2015}

\begin{subequations}\label{eq:cond_eq}
\begin{gather}
     d\mathbf{X}_c(t) =\mathbf{A} \mathbf{X}_c(t)dt + \mathbf{B}\mathbf{u}(t)dt \notag\\ + \left(\SigmaCon(t)\mathbf{C}^T+\mathbf{M}\right)  \mathbf{W}^{-1}d\mathbf{w}(t)\label{eq:cond_mean}\\
    d\SigmaCon = \mathbf{A}\SigmaCon(t)dt + \SigmaCon(t)\mathbf{A}^T dt+ \mathbf{V}dt\notag\\- \left(\SigmaCon(t)\mathbf{C}^T+\mathbf{M}\right)\mathbf{W}^{-1}\left(\SigmaCon(t)\mathbf{C}^T+\mathbf{M}\right)^Tdt.\label{eq:cond_cov}
   \end{gather}
\end{subequations}

From these equations it is evident that the dynamics of the conditional mean values are coupled to the second moments. They are driven by the feedback through $\mathbf{B}\mathbf{u}(t)$, as well as the  stochastic vector $\mathbf{dw}(t)$, such that $\mathbf{X}_c$ will follow a random path in phase space. In contrast, the second moments are independent from the mean values and not influenced by any feedback. They are completely determined by the balance between measurement and back-action noise on the one side and decoherence on the other. In what follows, we  focus on the normal-modes dynamics.

\subsection{Entanglement criteria}
In this work, we are interested in the bipartite entanglement of motion shared between the two masses. Due to the Gaussian character of the system, all entanglement properties are fully described by the system's $4\times 4$ (conditional) covariance matrix, which can be represented in its block-diagonal form using the $2\times2$ covariance matrices $\Sigma^\pm$ of the respective normal modes
\begin{equation}
     \Sigma= \begin{pmatrix}
        \Sigma^{+} & 0 \\
         0  & \Sigma^{-}
    \end{pmatrix}.
\end{equation}
For a bipartite system of $1\times 1$ bosonic modes the logarithmic negativity  $ \mathcal{E}_\mathcal{N} = \text{max}\{0, -\text{ln} (2\tilde{\nu})\}$ with symplectic eigenvalue $\tilde{\nu}$ is a necessary and sufficient measure of entanglement \cite{Adesso_2007, Werner_2002}
\begin{subequations}\label{eq:logNeg}
\begin{gather}
     \tilde{\nu} = \frac{1}{\sqrt{2}}\sqrt{\sigma-\sqrt{\sigma^2-4 \det\Sigma_+ \det\Sigma_-}}\\
     \sigma = \frac{\alpha_-^2}{\alpha_+^2}\Sigma^+_{11}\Sigma^-_{22}-2\Sigma^+_{12}\Sigma^-_{12}+\frac{\alpha_+^2}{\alpha_-^2}\Sigma^+_{22}\Sigma^-_{11}.
\end{gather}
\end{subequations}
The state is detected as entangled for $ \mathcal{E}_\mathcal{N} >0$, while all separable states yield $\mathcal{E}_\mathcal{N}=0$.

In addition to logarithmic negativity, we consider entanglement criteria that are linearly dependent on the quadrature covariances. Criteria of this kind have important practical implications as they can be realized experimentally \cite{Audenaert_2006, Guhne_2007}. In \sref{sec:LQR} we proceed with the analysis of a special class of entanglement criteria based on EPR variances. Using the definition $\Delta(\hat{O}):=  \langle \hat{O}^2 \rangle -\langle \hat{O}\rangle^2$ for the variance of an operator $\hat{O}$, we obtain the EPR-variance 
\cite{Duan_2000, Simon_2000}:
\begin{equation}\label{eq:fullEPR}
\begin{gathered}
       \Delta_{\rm EPR}:= \Delta\left(\hat{x}_1 + \cos\theta \hat{x}_2
       + \sin\theta \hat{p}_2\right)\\+\Delta\left(\hat{p}_1 +\sin\theta \hat{x}_2 - \cos\theta \hat{p}_2\right)
\end{gathered}    
\end{equation}
 which detects entanglement for $\Delta_{ \rm EPR}<2$. However, this criterion is only a sufficient and, generally, cannot detect all bipartite entangled Gaussian states, in contrast to the logarithmic negativity.

\section{Optimal Quantum Control}\label{sec:QuantumControl}
The estimation of the actual state of the system as well as its control can be formulated as a linear-quadratic-Gaussian (LQG) control problem \cite{Edwards_2005}.
Within this framework, the control algorithm aims to minimise a cost function while taking into account the noise acting on the system, as well as uncertainties linked to the measurement. 
As an important result of control theory, the separation principle states that the optimal control problem can be decomposed into two independent sub-problems: the design of an optimal observer (i) and an optimal regulator (ii). Both ensure stable dynamics of the system under control~\cite{Bouten_2005}. 
In this section, we apply this formalism to explore potential optimal control strategies to maximise the likelihood of entanglement generation.

\subsection{Kalman Filter}\label{sec:Kalman}
The Kalman filter estimates the state 
of a systems as formulated for the normal modes in \eref{eq:cond_eq}. This is accomplished iteratively by refining its state estimates through a recursive process. The procedure considers the full system dynamics, the statistical properties of the noise involved, and the continuous measurement with analysis of its outcome updated at the each time step \cite{Bechhoefer_2021}. Through this iterative mechanism, the Kalman filter optimally balances information from measurements and predictions, aiming to minimize the mean-squared estimation error. 

With the evolution of the conditional first and second moments presented in \eref{eq:cond_mean} and (\ref{eq:cond_cov}), the optimal quantum filter is formally equivalent to the \textit{classical Kalman-Bucy filter} \cite{Edwards_2005} and, therefore, we can build on principles of classical control theory \cite{Kalman_1961}. In this framework, it is standard practice to introduce the real $4\times2$ Kalman gain matrix  $\left(\SigmaCon(t) \mathbf{C}^T+\mathbf{M}\right)\mathbf{W}^{-1}$. Since we are focused on the stationary solution in this work, hereinafter, we assume that both normal modes have reached their steady state, such that $\dot{\Sigma}^c(t)=0$. Given this, the Kalman gain matrix becomes constant in time, and \eref{eq:cond_cov} transforms into the time-independent algebraic Riccati equation. With our choice of the measurement setup presented in \sref{sec:normal_modes}, the system of both normal modes is considered \textit{observable} in terms of control theory. This means that in principle, the Kalman filter can observe every state of the system  (see Appendix \ref{appendix:Obersvability_Controllability}). 

\subsection{Linear quadratic regulator (LQR)}\label{sec:LQR}
 The linear quadratic regulator (LQR) is a control strategy used to optimise the performance of a linear system by minimizing a quadratic cost function. It operates based on the knowledge provided by the Kalman filter and uses the conditional state to determine the optimal control input that minimizes the desired 
 combination of covariances. 
With the  real $4\times 4$ symmetric cost matrix $\mathbf{P}\geq0$ and the control effort matrix $\mathbf{Q}>0$, we consider the class of quadratic cost functions
\begin{equation}\label{eq:cost_func}
    \mathcal{J}= \int_{t_0}^{t_1}dt\left( \mathbb{E}[\mathbf{X}_c(t)^T \mathbf{P}\mathbf{X}_c(t) +  \mathbf{u}(t)^T \mathbf{Q}\mathbf{u}(t)] \right).
\end{equation}
Both, $\mathbf{P}$ and $\mathbf{Q}$ depend on the choice of the desired  state and the feedback setup. Since we are interested in the steady state of the system, we can formally extend the integration limit to infinite times.

The form of the cost matrix $\mathbf{P}$ depends on the actual performance objective, such as minimisation of the overall system energy (\textit{cooling}) \cite{Magrini_2021} or mechanical squeezing \cite{Isaksen_2023}.
Here, we focus on two cost matrices: 1 - $\mathbf{P}_{\rm cool}$ minimising the total energy of both normal modes, and 2 - $\mathbf{P}_{\rm EPR}$, which minimises the EPR-variance \cite{Mancini_2007}: 

\begin{subequations}\label{eq:cost-matrices}
    \begin{gather}
        \mathbf{P}_{\rm cool} = \text{diag}\left(\Omega_+, \Omega_+, \Omega_-, \Omega_-\right)\label{eq:Pcool}\\
        \PEPR(\theta) = \Omega_0\begin{pmatrix}
           \mathbf{P}_+(\theta) &0\\
            0 & \mathbf{P}_-(\theta+\pi) 
        \end{pmatrix} \\
        \mathbf{P}_\pm(\theta) = \begin{pmatrix}
           \frac{1+\cos \theta}{\alpha_\pm}&\sin \theta\\
            \sin\theta & \alpha_\pm^2(1-\cos \theta) 
        \end{pmatrix}.\label{eq:Pepr} 
    \end{gather}
\end{subequations}
The matrix \eref{eq:Pepr} is obtained from the EPR-variance \eref{eq:fullEPR}, taking into account the symplectic transformation \eref{eq:transform}.

The matrix $\mathbf{Q}$  depends on the actual feedback configuration. For the ensuing discussion, we focus on two feedback strategies: 1 - both trapped particles are controlled through the same feedback, and 2 - each particle is driven by an independent feedback. The first scenario could be achieved by applying the same time-dependent electric field $E_{\rm fb}(t)$ to both particles. To ensure the system is \textit{controllable}, we have to choose the excess charges on both particles to fulfil $|Q_1|\neq|Q_2|$, such that, the LQR is capable to control each state of the system within the given feedback-configuration (see Appendix \ref{appendix:Obersvability_Controllability}). This condition is typically met in current experiments that demonstrate Coulomb coupling between levitated particles~\cite{Rieser_2022,penny2023sympathetic,bykov20233d, Deplano_2024}. Given this, we can rewrite the control input $\mathbf{B}\mathbf{u}(t) =   \mathbf{B}_{\rm sin}u_{\rm sin}(t)$ using the real $4\times 1$   matrix $\mathbf{B}_{\rm sin}$ and the time-dependent scalar $u_{\rm sin}(t)$, where
\begin{subequations}
\begin{gather}
      \mathbf{B}_{\rm sin}= \begin{pmatrix}
        0 & \frac{1}{\alpha_+} & 0 & \frac{1}{\alpha_-}\frac{Q_1-Q_2}{Q_1+Q_2}
    \end{pmatrix}^T \\
    u_{\rm sin}(t) = (Q_1+Q_2)\frac{x_{\rm zpf}E_{\rm fb}(t)}{\sqrt{2}}\label{eq:sin_contr}
\end{gather}
\end{subequations}
The corresponding matrix for the control effort becomes a scalar $Q_{\rm sin}=q/\Omega_0 $. It quantifies the penalty on the control input and ensures the feedback force remains finite.

In the second scenario, we allow individual control of each particle position and/or frequency, for example, by configurable optical trap arrays~\cite{Vijayan_2023,Reisenbauer_2024}. Here, we use the control effort matrix $\mathbf{Q}_{\rm ind} = \text{diag}(q/\Omega_0, q/\Omega_0)$ to penalize both control inputs equally, with the same effort $q$. The feedback strategy based on two independent input forces, always ensures the system remains controllable.

The optimal state-based feedback is determined by the real $2\times 4$ feedback gain matrix $\mathbf{K}(t) = \mathbf{Q}^{-1}\mathbf{B}^T\Omega(t)$ and the conditional means  \cite{Edwards_2005}
\begin{equation}
    \mathbf{u}(t) = - \mathbf{K}(t) \mathbf{X}_c(t)
\end{equation}
where the time-evolution for $\Omega(t)$ is given by the Riccati-equation \cite{Edwards_2005, Bechhoefer_2021}, running backwards in time:

\begin{equation}\label{eq:feedback_gain}
\begin{gathered}
       - d\Omega(t)= \mathbf{A}^T \Omega(t)dt +\Omega(t) \mathbf{A}dt + \mathbf{P}dt \\
       - \Omega(t)  \mathbf{B} \mathbf{Q}^{-1} \mathbf{B}^T\Omega(t)dt.
\end{gathered}
\end{equation}
The time evolution can be rigorously derived by the variational calculus following Refs. \cite{Edwards_2005, Bechhoefer_2021}.  

Based on the conditional state obtained by the Kalman filter, the LQR drives the system into the  \textit{unconditional state} with density matrix $\hat{\rho} = \mathbb{E}[\hat{\rho}_c]$. This state is represented by the covariance matrix $\SigmaUn= \SigmaCon +\XiEx$, with positive definite  excess noise matrix $\XiEx=\mathbb{E}[\mathbf{X}_c \otimes \mathbf{X}_c^T]$ \cite{Hofer_2015}, that evolve in time evolution according to (see Appendix \ref{appendix:Excess_noise})

\begin{equation}\label{eq:Excess_noise}
\begin{gathered}
     d\XiEx(t)= \left(\mathbf{A}-\mathbf{B}\mathbf{K}(t)\right)\XiEx(t)dt +\XiEx(t) \left(\mathbf{A}-\mathbf{B}\mathbf{K}(t)\right)^Tdt \\
     + \left(\SigmaCon(t) \mathbf{C}^T+\mathbf{M}\right)\mathbf{W}^{-1}\left(\SigmaCon(t) \mathbf{C}^T+\mathbf{M}\right)^Tdt. 
\end{gathered} 
\end{equation}
Note, the excess noise results from the LQR relying on the system state estimation by the Kalman filter, leading to the feedback imperfections \cite{Isaksen_2023}. 

Due to the aforementioned seperation principle \cite{Bouten_2005}, the Kalman gain and feedback gain matrices remain independent on the LQR and Kalman filter, respectively. However, since the excess noise is influenced by both of them, as indicated in \eref{eq:Excess_noise},  the unconditional state strongly depends on the measurement setup, the feedback strategy and the cost funtion of LQG. Consequently, the conditional state always exhibits larger logarithmic negativity, serving as a fundamental limit for unconditional entanglement generation. An ideal cost function and feedback strategy should aim to minimize the difference between the unconditional and conditional covariances by reducing the excess noise.

\section{Stationary Gaussian entanglement}\label{sec:Results}

\subsection{Fundamental limit}\label{sec:Results_Conditional}

\begin{figure}[t]
    \centering
    \includegraphics[width=\columnwidth]{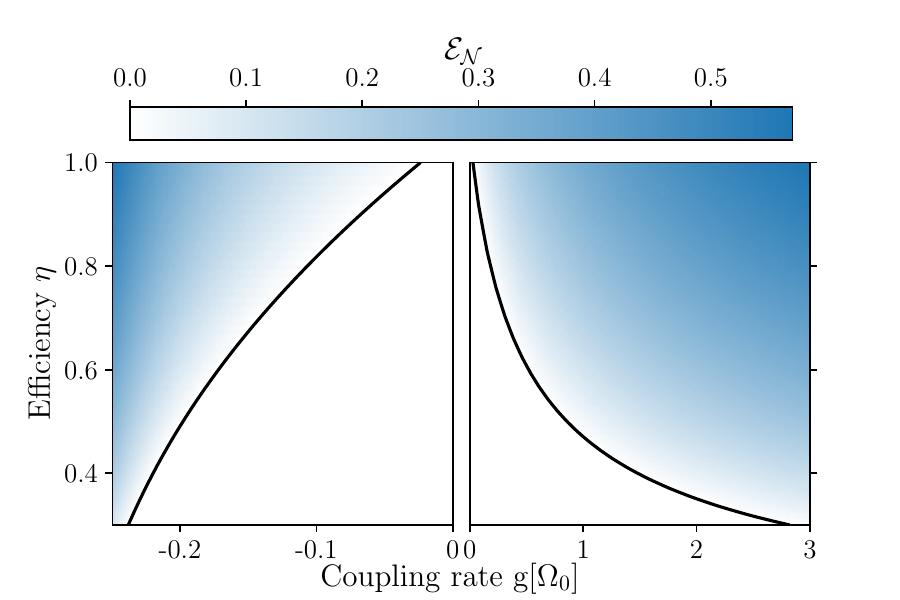}
    \caption{\label{fig:Fundamental limit}Steady-state conditional entanglement, quantified by the logarithmic negativity  depicted as a function of coupling rate $g$ and  efficiency $\eta$ for repulsive (left) and attractive (right) interactions. For this figure we set $\Gamma_{\rm ba}/\Omega_0 = 5\%$, $\Gamma_{\rm th}/\Gamma_{\rm ba} = 5\%$, and a quality factor $Q =\Omega_0/\gamma = 10^{10}$. The black line represents the boundary between separable ( $\mathcal{E}_\mathcal{N}=0)$ and entangled states ($ \mathcal{E}_\mathcal{N}>0)$.}
\end{figure}

The optimal estimation problem can be solved separately for each normal mode, when they are independently observable, as described in \sref{sec:Kalman}. We  consider the steady-state solution of $\SigmaCon$ for \textit{attractive} interaction in the limit of low damping and decoherence rates of the mechanical oscillators $\gamma \ll \Gamma_{\rm ba}, \Gamma_{\rm th} \ll \Omega_0 $. The correpsonding covariances and their approximated values are stated in Appendix \ref{appendix:Conditional_state}. 
In the case of low total decoherence rate $ \Gamma_{\rm th} +\Gamma_{\rm ba}\ll \Omega_0$, correlations between momentum and position of the normal modes vanish as $\left(\Gamma_{\rm ba}+\Gamma_{\rm th}\right)/\Omega_\pm$ decreases. The logarithmic negativity is given by

\begin{equation}
    \mathcal{E}_\mathcal{N} = -\frac{1}{2} \text{ln}\left(\frac{1}{\alpha_-^2}\frac{ \Gamma_{\rm th} +\Gamma_{\rm ba}}{\Gamma_{\rm m}}\right).
\end{equation}
This result yields a threshold for the coupling rate $g_+$ corresponding to the attractive interaction that enable entanglement in the conditional state. Given the prefect detection efficiency $\eta = 1$, i.e. $\Gamma_{\rm ba}=\Gamma_{\rm m}$, we arrive at the following requirement
\begin{equation}
    \frac{g_+}{\Omega_0}>\frac{1}{4}\left[\left(\frac{\Gamma_{\rm ba}+\Gamma_{\rm th}}{\Gamma_{\rm ba}}\right)^2-1\right].
\end{equation}
Alternatively, we can get similar condition for the detection efficiency, given that the coupling rate is above the threshold

\begin{equation}
    \eta_+>\frac{\Gamma_{\rm ba}+\Gamma_{\rm th}}{\Gamma_{\rm ba}}\frac{1}{\sqrt{1+4\frac{g}{\Omega_0}}}. 
\end{equation}
A low detection efficiency can, therefore, be compensated by stronger interparticle interaction. However, observing conditional entanglement below the critical coupling $g_+$ is not possible, regardless of the efficiency.

For \textit{repulsive} interaction ($g<0$),  we note that the differential mode frequency may, in principle, decrease below the typical decoherence rates $\Omega_-\ll \Gamma_{\rm ba},\Gamma_{\rm th}$ when $\alpha_-\ll 1$. In this case, the position variance decreases, while the momentum variance increases accompanied by the enhancement in correlations between momentum and position quadratures. As a result, an approximate expression for logarithmic negativity, similar to the one obtained above for the attractive interaction, no longer adequately describes entanglement. Therefore, we limit our analysis to the regime of $\alpha_-^4 > 2\sqrt{2\eta}\left(\Gamma_{\rm ba}+\Gamma_{\rm th}\right)/\Omega_0$, given the coupling rates $|g|/\Omega_0\gg 4-8\sqrt{2}\left(\Gamma_{\rm ba}+\Gamma_{\rm th}\right)/\Omega_0$, 
where the reduced expression for the logarithmic negativity remains a valid approximation.   
Under these assumptions, we can neglect contributions from position-momentum correlators $\SigmaCon_{x_-p_-}$, and arrive at a simple form of logarithmic negativity (see Appendix \ref{appendix:Conditional_state}).

\begin{equation}
     \mathcal{E}_\mathcal{N} = -\frac{1}{2} \text{ln}\left(\alpha_-^2\frac{\Gamma_{\rm ba}+\Gamma_{\rm th}}{\Gamma_{\rm m}}\right). 
\end{equation}
Similar analysis identifies a threshold of the interaction strength  $g_-$ for the fundamental limit of the entanglement generation. 
We find the entanglement thresholds in terms of the coupling rate $|g| > |g_-|$ and detection efficiency $\eta_-$:
\begin{subequations}
    \begin{gather}
        \frac{g_-}{\Omega_0} < \frac{1}{4}\left[\left(\frac{\Gamma_{\rm ba}}{\Gamma_{\rm ba}+\Gamma_{\rm th}}\right)^2-1\right]\\
        \eta_- > \frac{\Gamma_{\rm ba}+\Gamma_{\rm th}}{\Gamma_{\rm ba}}\sqrt{1+\frac{4g}{\Omega_0}}. 
    \end{gather}
\end{subequations}
Unlike the attractive regime of interaction, here one can compensate imperfection in the detection scheme by the interaction strength up to the level of $|g|/\Omega_0 < 0.25$ .

When the back-action noise dominates over the thermal noise, e.g. for optically trapped solids in ultra-high vacuum~\cite{Magrini_2021, Tebbenjohanns_2021}, the total decoherence can be well approximated by the back-action mechanism $\left(\Gamma_{\rm ba}+\Gamma_{\rm th}\right)\approx\Gamma_{\rm ba}$.
In this limit, we find the entanglement threshold conditions for attractive interaction $g_+/\Omega_0 >0$ and $\eta_+>1/\sqrt{1+4g/\Omega_0}$, which represents the fundamental limit of entanglement generation. Achieving such ultra-strong interaction between two masses (e.g. $g \sim \Omega_0$ for $\eta \sim 0.5$)  poses significant experimental challenges. In contrast, the entanglement under repulsive interaction can be achieved at much lower coupling rates $|g_-|/\Omega_0 >0$ and efficiencies $\eta_- > \sqrt{1+4g/\Omega_0}$. In particular, the required conditions are  met for e.g. $g/\Omega_0\sim -0.18$ at $\eta\sim 0.5$. \fref{fig:Fundamental limit} shows logarithmic negativity of the conditional state as function of the interaction strength $g/\Omega_0$ and detection efficiency $\eta$, for the repulsive (left) and attractive (right) configurations. We assume $\Gamma_{\rm ba}/\Omega_0 = 5\%$ and $\Gamma_{\rm th}/\Gamma_{\rm ba} = 5\%$, which is in a good agreement with recent experiments \cite{Tebbenjohanns_2021}. The difference in performance between the two regimes can be understood by analyzing the equations for reduced logarithmic negativity, Eq. (20) and Eq. (23), given the amplitude $\alpha_-$ at the normal mode transformation \eref{eq:transform} expressed in terms of interaction strength $|g|$. Specifically, $\mathcal{E}_\mathcal{N}$ scales as $\sqrt{1 + 4|g|/\Omega_0}$ for the attractive interaction, and as ${\sqrt{1 - 4|g|/\Omega_0}}^{-1}$ for the repulsive interaction. Obviously, for the repulsive regime, logarithmic negativity grows much faster with interaction. The corresponding measurement and decoherence rate increase as well with the coupling rate, as follows from the matrices $\mathbf{C}$ (\ref{eq:measurement-mat}) and $\mathbf{V}$ (\ref{eq:Vmat}). However, the ratio between the total decoherence and the measurement rate remains constant. As a result, the evolution of the differential mode, slowed down by a factor of $\alpha_-^2$, is effectively measured at a rate amplified by a factor of $\sim {\sqrt{1 - 4|g|/\Omega_0}}^{-1}$, maintaining the same information loss. This enhances squeezing in the position-momentum quadrature of the differential mode, which is crucial for the fundamental limit of entanglement generation, see \eref{eq:logNeg}. Therefore, we believe that the repulsive configuration is the most promising approach for experimental implementation.

\subsection{Unconditional entanglement generation}
\begin{figure}[t]
\includegraphics[width=\columnwidth]{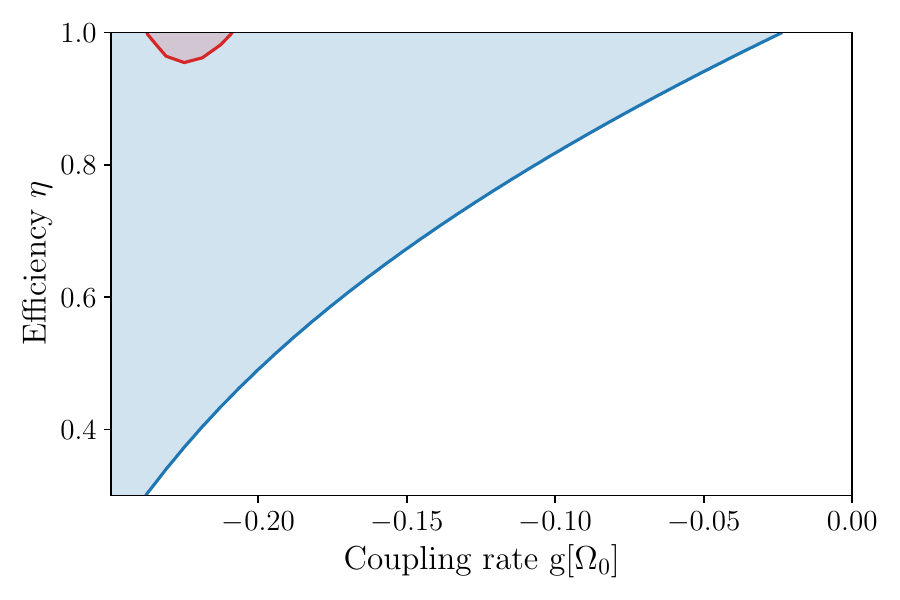}
\caption{\label{fig:Uncond_Cost} Steady-state separability bound for conditional and unconditional entanglement for repulsive interaction with $\PEPR$ as cost matrix as a function of coupling rate $g$ and detection efficiency $\eta$. The blue area represents states which are conditional entangled while for the red area we additionally observe unconditional entanglement. For the single feedback we choose $Q_1/Q_2 = 5$ and the control effort $q= 0.1$. All other parameters are the same as in \fref{fig:Fundamental limit}. For a cost matrix $\mathbf{P}_{\mathrm{cool}}$ simulations do not yield entanglement in this parameter regime.} 
\end{figure}

We first compare the two cost functions associated with $\mathbf{P}_{\rm cool}$ and $\PEPR$ presented in \eref{eq:cost-matrices}. 
Assuming both particles are controlled by the same input  $u_{\rm sin}$ from \eref{eq:sin_contr} 
we chose $|Q_1|/|Q_2|= 5$ and control effort $q=0.1$, both within the reach of current experiments \cite{Magrini_2021}. 
To maximize unconditional entanglement in the system we choose $\theta=\pi$ ($\theta=0$) being optimal for attractive (repulsive) interaction. Importantly, the system of two particles is both \textit{observable} and \textit{controllable}, ensuring that a solution for the unconditional state with the single feedback strategy exists \cite{Bittanti_2012}.
In \fref{fig:Uncond_Cost} we show the fundamental limit of entanglement generation represented by the conditional state (blue), and the separability bound for the entanglement in unconditional states obtained by the \textit{EPR} cost function (red). For consistency, we use the same parameters as in \fref{fig:Fundamental limit}, namely $\Gamma_{\rm ba}/\Omega_0 = 5\%$, $\Gamma_{\rm th}/\Gamma_{\rm ba} = 5\%$, and $Q =\Omega_0/\gamma = 10^{10}$. Note that in the case of repulsive interaction with the \textit{cooling} cost function, the system under the single feedback does not demonstrate entanglement in the unconditional state, regardless of the interaction strength and detection efficiency. This is in agreement with recent theoretical reports \cite{Rudolph_2022}

\begin{figure}[t]
\includegraphics[width=\columnwidth]{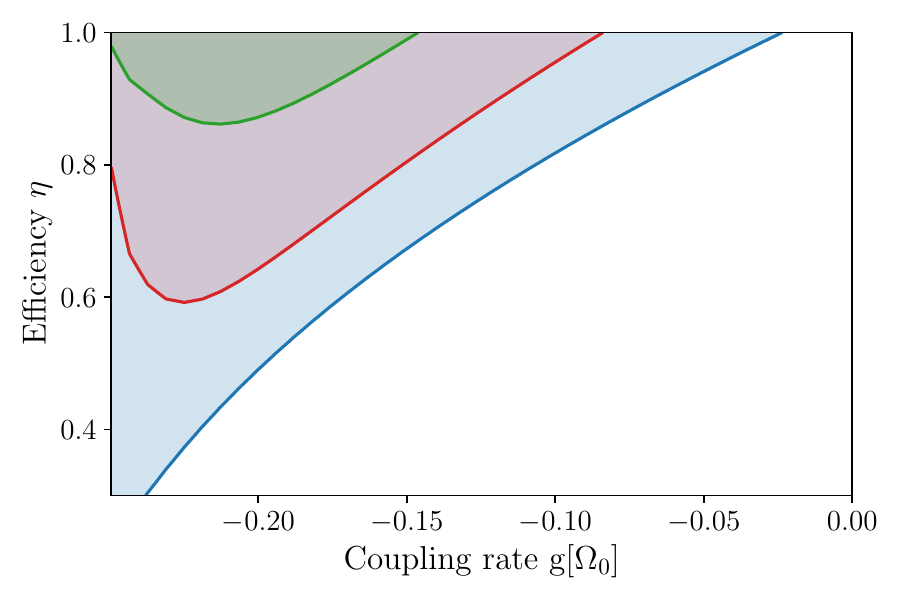}
\caption{\label{fig:Uncond_SingInd} Steady-state separability bound for conditional  and unconditional entanglement for repulsive interaction with  EPR-  and cooling-cost matrix applying individual feedback as discussed in \sref{sec:LQR}. The blue area represents conditional entangled states, the red and green area correspond to unconditional entanglement for EPR- and cooling-cost. The parameters are the same as in \fref{fig:Fundamental limit} and \ref{fig:Uncond_Cost}.} 
\end{figure}

In contrast to the single feedback strategy, the independent control of each particle makes the LQR problem essentially separable for both normal modes. Applied to the entanglement problem, it results in significantly lower separability bounds for the unconditional state, as shown in \fref{fig:Uncond_SingInd}. Although the \textit{cooling} cost function demonstrates unconditional entanglement (green) under extremely stringent conditions, the \textit{EPR}-cost performs much better. Even for an asymptotic feedback $q\rightarrow 0$, the \textit{cooling}-cost adds additional noise to the unconditional state on the order of $\Gamma_{\rm m}/\Omega_0$~\cite{Isaksen_2023}, which suppresses entanglement in the system. The problem was previously identified in \cite{Rudolph_2022}, where authors attribute the lack of unconditional entanglement to the excess noise from the filter. In contrast, \textit{EPR} cost function under the asymptotic feedback can reach fundamental limit of entanglement in the unconditional state. This effect is independent of the specific interaction, whether it is attractive or repulsive. However, at the finite $q>0$, there is an excess noise, that elevates separability bounds for unconditional state comparing the conditional one, see \fref{fig:Uncond_SingInd}. With the control matrix for the independent feedback \eref{eq:control_matrix} we can find closed analytical solutions of the steady state excess-noise \eref{eq:Excess_noise} and, thus, the unconditional state, for both \textit{cooling} and \textit{EPR} cost functions \cite{Bittanti_2012}. 
The corresponding covariances of the unconditional state in the limit $\gamma\ll\Omega_0$ are given in appendix \ref{appendix:Uncond_indv}.

We outline and discuss the main results of this section: 

(i) -- The choice of cost functions significantly impacts unconditional entanglement generation. The \textit{EPR} cost function consistently outperforms the \textit{cooling} cost function in both single and independent feedback scenarios. This can be interpreted based on the nature of the cost matrices: the \textit{cooling} cost function minimizes the total energy of the normal modes, whereas the \textit{EPR} cost function enhances EPR-like squeezing between the particles, promoting entanglement generation. By reducing excess noise, the \textit{EPR} cost function results in strongly correlated motion of the particles.

(ii) -- Independent feedback strategy exhibits superior performance compared to the single feedback. Due to its low frequency, the differential mode under repulsive interaction is particularly sensitive to excess noise arising from estimation errors in the single feedback scenario. This leads to the inflation of the squeezed state of the differential mode in phase space, compromising unconditional entanglement.

(iii) -- The conditional quantum state employed as the optimal estimator for LQR state-based feedback, combined with an \textit{EPR} cost function allows for steady-state unconditional entanglement in the repulsive regime of electrostatic interactions \(|g|/\Omega_0 \sim 0.2\).


While our results are, in principle, within experimental reach, several significant challenges remain on the way toward achieving unconditional steady-state entanglement.
Let’s consider optically levitated systems \cite{Delic_2020, Magrini_2021, Tebbenjohanns_2021} which are promising candidates due to their excellent environmental isolation and quantum control capabilities ~\cite{Ballestero_2021}.
For silica particles ($\rho = 1850 \rm~kg/m^3$, $\chi_e = 0.8$) with a radius of $R = 50 \rm~nm$ each, trapped in an optical tweezer with a wavelength of $\lambda = 1550 \rm ~nm$, trapping power of $P = 100 \rm ~mW$, and Rayleigh length of $x_R = 1.76 ~\text{µm}$, the resulting trapping frequency is $ \Omega_0/2\pi= 29.5 \rm ~kHz$ \cite{Isart_2019}.
Assuming the particles interact via the Coulomb force with charges $Q_{1,2} = 50 \rm e$, the coupling constant $C = Q_1 Q_2 / 4\pi\epsilon_0$ allows the critical coupling $g_{crit}^-$ to be reached at a separation distance of $d = 3.5 ~\text{µm}$.
The back-action rate in this case is $\Gamma_{\rm ba}/\Omega_0\approx 5\%$ \cite{Henning_2021_b}.
Operating at room temperature $T=300 \rm K$ under a nitrogen atmosphere at a pressure of $2\cdot10^{-10} \rm~ mBar$ yields  $\gamma/2\pi\approx 0.4  \text{µHz}$ and $\Gamma_{\rm th}/\Gamma_{\rm ba}\approx 5\%$ \cite{Magrini_2021}. To achieve unconditional steady-state entanglement in this setting requires detection efficiencies of at least $\eta\approx0.7$. High detection efficiency, along with the need for independent control and the low frequency of the differential mode, pose substantial experimental challenges.

Additionally, in practical applications, although LQG control benefits from knowledge of the system dynamics and model-based measurements, uncertainties in the model or its parameters, such as frequency shifts, can undermine the performance of LQG control \cite{Bechhoefer_2021}.

\section{Conclusion}

In conclusion, our study shows that a carefully chosen control objective and feedback strategy for LQG control expands the parameter space for  steady-state unconditional entanglement generation between two masses interacting via direct physical force, without the need for cavities to mediate their interaction. 
The implementation of an EPR-type minimisation constraint in the LQR consistently outperforms conventional cooling strategies. When combined with independent control of each mass motion, it enables unconditional entanglement for repulsive interactions, not accessible otherwise~\cite{Rudolph_2022}. 
As a result, our approach demonstrates the feasibility of achieving unconditional entanglement with an  interaction strength \(|g|/\Omega_0 \sim 0.2\), which is an order of magnitude lower than reported in \cite{Henning_2021_b}. 

Future research should aim to develop more effective cost functions, which introduce necessary and sufficient entanglement witnesses, as proposed in Ref. \cite{Duan_2000, Simon_2000}. Another promising approach employs time-dependent modulations of systems parameters, e.g. the interaction strength, which  results in nonequilibrium entanglement generation. In Ref. \cite{poddubny2024nonequilibrium} we address the problem within the optimal quantum control formalism, using the insights on control objectives and feedback strategies from this work.  
Our study paves the way for entanglement generation between large masses interacting directly via $1/r^{n}$ potentials under continuous, strong position measurement.


\begin{acknowledgments}

The authors thank Henning Rudolph, Ayub Khodaee, Nancy Gupta, Nikolai Kiesel and Corentin Gut for insightful  discussions. 
K.W. and M.A. received funding from the European Research Council (ERC) under the European Union’s Horizon 2020 research and innovation program (grant agreement No 951234), and from the Research Network Quantum Aspects of Spacetime (TURIS). K.W.  acknowledges the support by the Vienna Doctoral School in Physics (VDSP). B.A.S. acknowledges funding by the DFG–510794108 as well as by the Carl-Zeiss-Foundation through the project QPhoton. U.D. acknowledges funding from the Austrian Science Fund (FWF, Project DOI 10.55776/I5111).
A.D. acknowledges partial support by the Austrian Science Fund (FWF) [10.55776/COE1, PAT9140723] and the European Union – NextGenerationEU. A.V.Z. acknowledges support from the European Union’s
Horizon 2020 research and innovation programme under the Marie Sklodowska-Curie grant LOREN (grant
agreement ID: 101030987).
\end{acknowledgments}

\appendix 

\section{Observability and Controllability}\label{appendix:Obersvability_Controllability}
For a detailed discussion on observability and controllability in LQG problems we refer the reader to the work in \cite{Wiseman_2009, Bechhoefer_2021}, here we give a short explanation and as mathematical formulation. 

\textit{Observability}  refers to the ability to determine the system's internal state based on its output measurements.  Mathematically, for a linear time-invariant system represented by the time evolution in \eref{eq:cond_mean} and (\ref{eq:cond_cov}), observability is related to the rank of the observability matrix 
\begin{equation}
    \mathcal{O} = \begin{bmatrix}\mathbf{C} \\ \mathbf{CA} \\ \mathbf{CA^2} \\ \vdots \\ \mathbf{CA^{n-1}} \end{bmatrix} \
\end{equation}

where $n$ refers to the dimension of the state space (here $n=4$). The system is said to be observable if the rank of $\mathcal{O}$ is $n$. Observability ensures that 
no state is completely hidden, or ``\textit{unobservable}" from the available measurements.

\textit{Controllability} is given whenever the input $\mathbf{u}(t)$ can manipulate the system from any initial state to any desired state or condition within a finite time. Mathematically, again using the time evolution from \eref{eq:cond_mean} and (\ref{eq:cond_cov}), we can define the controllability matrix $\mathcal{C}$ as

\begin{equation}
     \mathcal{C} = \begin{bmatrix} \mathbf{B} & \mathbf{AB} & \mathbf{A^2B} & \ldots & \mathbf{A^{n-1}B} \end{bmatrix}. 
\end{equation}

The system is said to be controllable if the rank of $\mathcal{C}$ is equal to $n$. This condition ensures that the control input $\mathbf{u}$ can reach and manipulate all possible states of the system. 

\section{Excess noise evolution}\label{appendix:Excess_noise}

A detailed derivation for the excess noise can be found in the appendix of \cite{Hofer_2015}, here we present the main results. For the conditional state the covariance matrix is defined by 

\begin{equation}
\begin{gathered}
    \hat{\Sigma}^c_{ij} = \frac{1}{2} \left(\langle \hat{X}_i \hat{X}_j\rangle_c +\langle \hat{X}_j \hat{X}_i\rangle_c\right)-\langle \hat{X}_i \rangle_c \langle \hat{X}_j \rangle_c\\
    = \mathrm{Re}\left(\langle \hat{X}_i \hat{X}_j\rangle_c\right) -\langle \hat{X}_i \rangle_c \langle \hat{X}_j \rangle_c
    \end{gathered}
\end{equation}
which can be compactly written using the vector notation introduced in \sref{sec:Sys_dynm}: 
\begin{equation}
     \hat{\Sigma}^c= \mathrm{Re}\left(\langle(\hat{\mathbf{X}} - \hat{\mathbf{X}}_c) (\hat{\mathbf{X}}- \hat{\mathbf{X}}_c)^T \rangle \right). 
\end{equation}
Using the fact that $(\hat{\mathbf{X}} - \hat{\mathbf{X}}_c^T) $ is uncorrelated with the conditional mean, i.e. $\langle(\hat{\mathbf{X}} - \hat{\mathbf{X}}_c)\hat{\mathbf{X}}_c\rangle=0$  from the orthogonality principle  \cite{Edwards_2003, Bouten_2007}, we find for the unconditional state where $\langle\hat{\mathbf{X}}\rangle$ is zero by construction

\begin{equation}
    \mathrm{Re}\left( \langle \hat{\mathbf{X}} \hat{\mathbf{X}}^T\rangle \right) = \SigmaCon + \XiEx
\end{equation}
with  the excess noise $\XiEx=\langle \hat{\mathbf{X}}_c \hat{\mathbf{X}}_c^T\rangle$. The corresponding equations of motion can be derived using \eref{eq:cond_mean}, for the steady-state we find

\begin{equation}
    0= \left(\mathbf{A}-\mathbf{B}\mathbf{K}\right)\XiEx +\XiEx\left(\mathbf{A}-\mathbf{B}\mathbf{K}\right)^T + \mathbf{k}\mathbf{W}\mathbf{k}^T
\end{equation}

\section{Conditional steady-state}\label{appendix:Conditional_state}
Solving the algebraic Riccati equation arising from the steady-state equations in \eref{eq:cond_cov} and introducing the total decoherence rate $\Gamma_{\rm tot}= \Gamma_{\rm ba}+\Gamma_{\rm th}$, we find:

\begin{subequations}
\begin{gather}
    \SigmaCon_{\rm x_sx_s}= -\frac{\alpha_s^2\gamma}{8\Gamma_{\rm m}}\notag\\
    + \frac{\alpha_s}{8\Gamma_{\rm m}}\sqrt{\alpha_s^2\gamma^2 + 2\Omega_s\left(\sqrt{16\Gamma_{\rm m}\Gamma_{\rm tot}+\alpha_s^4\Omega_s^2}-\alpha_s^2\Omega_s\right)}\\
    \SigmaCon_{ \rm x_sp_s} = \frac{4\Gamma_{\rm m}}{\alpha_s^2\Omega_s}\left(\SigmaCon_{\rm x_sx_s}\right)^2\\
     \SigmaCon_{\rm p_s p_s} = \left(\frac{\sqrt{16\Gamma_{\rm m}\Gamma_{\rm tot}+\alpha_s^4\Omega_s^2}}{\alpha_s^2\Omega_s}-\frac{4\Gamma_{\rm m}\gamma}{\alpha_s^2\Omega_s^2}\SigmaCon_{\rm x_sx_s}\right)\SigmaCon_{\rm x_sx_s}
\end{gather}
\end{subequations}
with $ s \in \{+, -\} $. 

In the limit of high quality factors $Q= \Omega_0/\gamma\gg 1$,  low decoherence rates $\Gamma_{\rm ba} , \Gamma_{\rm th} \ll \Omega_0$ and assuming  $\alpha_s^4 > \textcolor{red}{2}\sqrt{2\eta}\Gamma_{\rm tot}/\Omega_0$ the conditional covariance can be approximanted by

\begin{subequations}
\begin{gather}
       \SigmaCon_{ \rm x_sx_s} \approx \SigmaCon_{\rm p_s p_s} \approx \sqrt{\frac{\Gamma_{\rm tot}}{4\Gamma_{\rm m}}}\\
        \SigmaCon_{ \rm x_sp_s} \approx\frac{ \Gamma_{\rm tot}}{\alpha_s^2\Omega_s}.
\end{gather}
\end{subequations}

\section{Unconditional state for independent feedback}\label{appendix:Uncond_indv}


The unconditional covariance-matrix will be ``inflated" by the excess noise $\XiEx$ relative to the conditional covariances $\SigmaCon$. From the steady-state equations of motion in \eref{eq:Excess_noise} it is straightforward to prove $\SigmaUn_{x_sp_s} = 0$ for both normal modes, independent of the actual cost-function and cost-matrix. 
As stated above, for independent feedback the full LQG problem can be separated for both normal modes, allowing us to give a closed form for the unconditional covariance matrix. Since for this work we are interested in the limit of asymptotic feedback i.e. when the control effort approaches $q\rightarrow0$, we Taylor-expand the covaricane up to leading order in $q$ as proposed by \cite{Isaksen_2023}. 
Since $\SigmaUn_{x_sp_s} = 0$, the  logarithmic negativity of the unconditional state will depend on the product $\SigmaUn_{x_\pm x_\pm} \cdot\SigmaUn_{p_\mp p_\mp}$.

Employing a cooling-cost from\eref{eq:Pcool} onto the system and assuming the high-Q limit where $\gamma\ll \Omega_0$, we find for the unconditional covariances up to leading order in $q$: 

\begin{subequations}
    \begin{gather}
        \SigmaUn_{\rm x_sx_s} =\SigmaCon_{\rm x_sx_s} +\frac{4\Gamma_{\rm m}}{\alpha_s^2\Omega_s}(\SigmaCon_{\rm x_sx_s})^2  + \sqrt{q}\frac{4\Gamma_{\rm m}}{\alpha_s^2\sqrt{\Omega_0\Omega_s}}\SigmaCon_{\rm x_sx_s}(\SigmaCon_{\rm x_sx_s}+\SigmaCon_{ \rm x_sp_s})\notag\\
       +q\frac{4\Gamma_{\rm m}}{\alpha_s^2\Omega_0}\left(\left(\SigmaCon_{ \rm x_sp_s}\right)^2-3\left(\SigmaCon_{\rm x_sx_s}\right)^2\right) +\mathcal{O}(q^{3/2})\\
       \SigmaUn_{\rm p_sp_s} =\SigmaCon_{\rm p_s p_s} +\frac{4\Gamma_{\rm m}}{\alpha_s^2\Omega_s}(\SigmaCon_{\rm x_sx_s})^2  + \sqrt{q}\frac{4\Gamma_{\rm m}}{\alpha_s^2\sqrt{\Omega_0\Omega_s}}\left(\left(\SigmaCon_{ \rm x_sp_s}\right)^2-\left(\SigmaCon_{\rm x_sx_s}\right)^2\right)\notag\\
       -q\frac{4\Gamma_{\rm m}}{\alpha_s^2\Omega_0}\left(\left(\SigmaCon_{ \rm x_sp_s}\right)^2-3\left(\SigmaCon_{\rm x_sx_s}\right)^2\right)+\mathcal{O}(q^{3/2}).
    \end{gather}
\end{subequations}

 Considering an EPR-cost function as introduced in \eref{eq:Pepr} with $\theta=0$ and again in the high-Q limit we find
\begin{subequations}
\begin{gather}
        \SigmaUn_{\rm x_+x_+}= \SigmaCon_{\rm x_+x_+}+ q^{1/4}\frac{6\cdot2^{1/4}\Gamma_{\rm m}}{\Omega_0} \left(\SigmaCon_{\rm x_+x_+}\right)^2 + \sqrt{q}\frac{4\sqrt{2}\Gamma_{\rm m}}{\Omega_0} \SigmaCon_{\rm x_+x_+} \SigmaCon_{\rm x_+p_+}\notag\\
        +q^{3/4}\frac{\Gamma_{\rm m}}{2^{1/4}\Omega_0}\left( 2\left(\SigmaCon_{\rm x_+p_+}\right)^2- \left(\SigmaCon_{\rm x_+x_+}\right)^2\right)+ \mathcal{O}(q^{5/4})\\
        \SigmaUn_{\rm p_-p_-}=  \SigmaCon_{\rm p_-p_-} + \sqrt{q}\frac{2\sqrt{2}\Gamma_{\rm m}}{\alpha_-^3\Omega_0 }\left( \left(\SigmaCon_{\rm x_-p_-}\right)^2+\left(\SigmaCon_{\rm x_-x_-}\right)^2\right) + \mathcal{O}(q^{3/2}).
\end{gather}
\end{subequations}

With the EPR-cost function for $\theta=\pi$ with the same assumpions as before, we have
 \begin{subequations}
     \begin{gather}
         \SigmaUn_{\rm p_+p_+}=  \SigmaCon_{\rm p_+p_+} + \sqrt{q}\frac{2\sqrt{2}\Gamma_{\rm m}}{\Omega_0}+\left( \left(\SigmaCon_{\rm x_+p_+}\right)^2+ \left(\SigmaCon_{\rm x_+x_+}\right)^2\right) + \mathcal{O}(q^{3/2})\\
         \SigmaUn_{\rm x_-x_-}= \SigmaCon_{\rm x_-x_-}+ q^{1/4}\frac{6\cdot2^{1/4}\Gamma_{\rm m}}{\alpha_-^{5/2}\Omega_0} \left(\SigmaCon_{\rm x_-x_-}\right)^2+ \sqrt{q}\frac{4\sqrt{2}\Gamma_{\rm m}}{\alpha_-\Omega_0} \SigmaCon_{\rm x_-x_-} \SigmaCon_{\rm x_-p_-}\notag\\
          +q^{3/4}\frac{\sqrt{\alpha_-}\Gamma_{\rm m}}{2^{1/4}\Omega_0}\left( 2\left(\SigmaCon_{\rm x_-p_-}\right)^2- \left(\SigmaCon_{\rm x_-x_-}\right)^2\right)+ \mathcal{O}(q^{5/4})
     \end{gather}
 \end{subequations}

\bibliography{refs}

\end{document}